# Giant spin-vorticity coupling excited by shear-horizontal surface acoustic waves

Mingxian Huang, Wenbin Hu, Huaiwu Zhang, Feiming Bai[*]

State Key Laboratory of Electronic Thin Films and Integrated Devices, University of

Electronic Science and Technology of China, Chengdu 610054, China

* To whom correspondence should be addressed. Electronic mails:

fmbai@uestc.edu.cn

**Abstract**

A non-magnetic layer can inject spin-polarized currents into an adjacent ferromagnetic layer via spin vorticity coupling (SVC), inducing spin wave resonance (SWR). In this work, we present the theoretical model of SWR generated by shear-horizontal surface acoustic wave (SH-SAW) via SVC, which contains distinct vorticities from well-studied Rayleigh SAW. Both Rayleigh- and SH-SAW delay lines have been designed and fabricated with a $Ni_{81}Fe_{19}$/Cu bilayer integrated on ST-cut quartz. Given the same wavelength, the measured power absorption of SH-SAW is four orders of magnitudes higher than that of the Rayleigh SAW. In addition, a high-order frequency dependence of the SWR is observed in the SH-SAW, indicating SVC can be strong enough to compare with magnetoelastic coupling.

## I. Introduction

Spin wave (SW) is the collective motion of magnons in a spin system of ordered magnetism. Due to its low power consumption [1], short wave length[2-4] and special phase property, spin wave can be used in a wide variety fields, such as information processing[5-7] , signal transmission[8-10] and logic devices [11,12]. In ferromagnets (FM), SWs typically have frequencies in the ~GHz range, with wavelengths ranging from hundreds of nanometers (dominated by quantum exchanges) to a few microns (dominated by dipole interactions). Generally, SWs are excited by spatial non-uniform alternating magnetic fields using antennas or transmission lines[13]. However, generating SWs with high amplitudes is challenging due to the mismatch between the SW wavelength and that of the electromagnetic waves (EMWs). Moreover, the decay rate of spin wave in most magnetic materials is extremely fast due to the damping of magnetic precession, which also limits the propagation distance of SWs. Therefore, how to realize effective excitation of SWs has been the focus of researchers for a long time.

Magnon-phonon coupling is an emerging means to excite and control spin waves. Unlike EMWs, surface acoustic waves (SAWs) can travel as far as millimeters in piezoelectric crystals, and have orders of magnitude smaller wavelengths that can match with SWs[14,15]. Recently, it has been demonstrated that a Rayleigh-mode surface acoustic wave (R-SAW) could be used to excite   spin wave resonance (SWR) over long distances with low power [16-25]. Conversion between R-SAW and SW has been directly visualized by Blai Casals et al[16].

There are two major ways of utilizing SAWs to excite SWs, one is magnetoelastic coupling (MEC) [16-22], the other is spin-vorticity coupling (SVC) [23-32]. For MEC, FM materials with a large magnetostriction coefficient is needed, but such materials are typically accompanied by large damping factors, which is not desired for SW transmission. For SVC, it can accumulate spin polarization in a non-magnetic (NM) metal, especially with a long spin lifetime or a weak spin-orbit coupling [27]. The diffusion of this spin accumulation to an adjacent FM layer can excite SWs in a FM/NM structure by the spin transfer torque (STT) [24]. Apparently, this provides a high degree of freedom in the selection of magnetic materials, so that materials with zero magnetostriction and a very low damping factor, such as permalloy ($Ni_{81}Fe_{19}$), can be chosen. However, experimental results reported so far show that the SWR excited by SVC is much weaker than that of MEC [24][30].

Therefore, it is critical to increase the SAW-to-SW conversion efficiency of SVC. Different from EMWs excitation, the diversity of SAW modes provides abundant opportunities for exploring various types of SVC. In current work, we will demonstrate theoretically and experimentally that a shear horizontal (SH)-mode SAW (SH-SAW) can effectively excite SWR in a $Ni_{81}Fe_{19}$/Cu structure via SVC, which is four orders of magnitudes stronger than that by a Rayleigh-mode SAW. The strong SWR can be attributed to the high frequency brought by the high phase velocity together with the large in-plane effective driven field generated by the out-of-plane vorticity component of the SH-SAW.

The paper is structured as follows. In Sec. II, the vorticities of R- and SH-SAW were deduced and compared, which were also verified by FEM eigenfrequency simulations. Then the spin-polarized currents (Sec. II A) and the effective driven fields (Sec. II B) due to the STT effect were calculated. The energy dissipation from SWR or the absorption of SAW power was then obtained based on the effective driven field and the magnetic susceptibility (Sec. II C). The details of device fabrication and measurement setup are present in Sec. III. Sec. IV is devoted to discuss the measured power absorption of two types of SAWs and the frequency dependence of SVC in the SH-SAW.

## II. Theory

In this section, we provide the theoretical framework for studying SWRs in a FM/NM structure. Firstly, the spin-polarized currents excited by SH- and R-SAWs via SVC are derived and compared. Then, the effective driven field generated by the spin current injection is given based on the STT analysis. Finally, and the power absorption originated from SWRs is calculated using the Landau–Lifshitz–Gilbert (LLG) equation, which also provides a foundation for our experimental work.

### A. Spin Current Generated by the SH- and R-SAWs Via SVC

The Hamiltonian for the SVC is defined as[24][27]:

$$\boldsymbol{H}_S = -\frac{\hbar}{2}\boldsymbol{\sigma}\cdot\boldsymbol{\Omega} \quad (1)$$

where $\hbar$ is the reduced Planck constant and $\boldsymbol{\sigma}$ is the Pauli matrix. $\boldsymbol{\Omega}$ represents the macroscopic mechanical rotational motion of the lattice, which can be described by the lattice displacement vector $\boldsymbol{u}$:

$$\boldsymbol{\Omega} = \frac{1}{2}\nabla\times(\partial\boldsymbol{u}/\partial t) == \begin{vmatrix} \vec{x} & \vec{y} & \vec{z} \\ \frac{\partial}{\partial x} & \frac{\partial}{\partial y} & \frac{\partial}{\partial z} \\ \frac{\partial u_1}{\partial t} & \frac{\partial u_2}{\partial t} & \frac{\partial u_3}{\partial t} \end{vmatrix} \quad (2)$$

$u_i(i=1,2,3)$ represent the displacements along the *x, y*, and *z* directions, respectively.

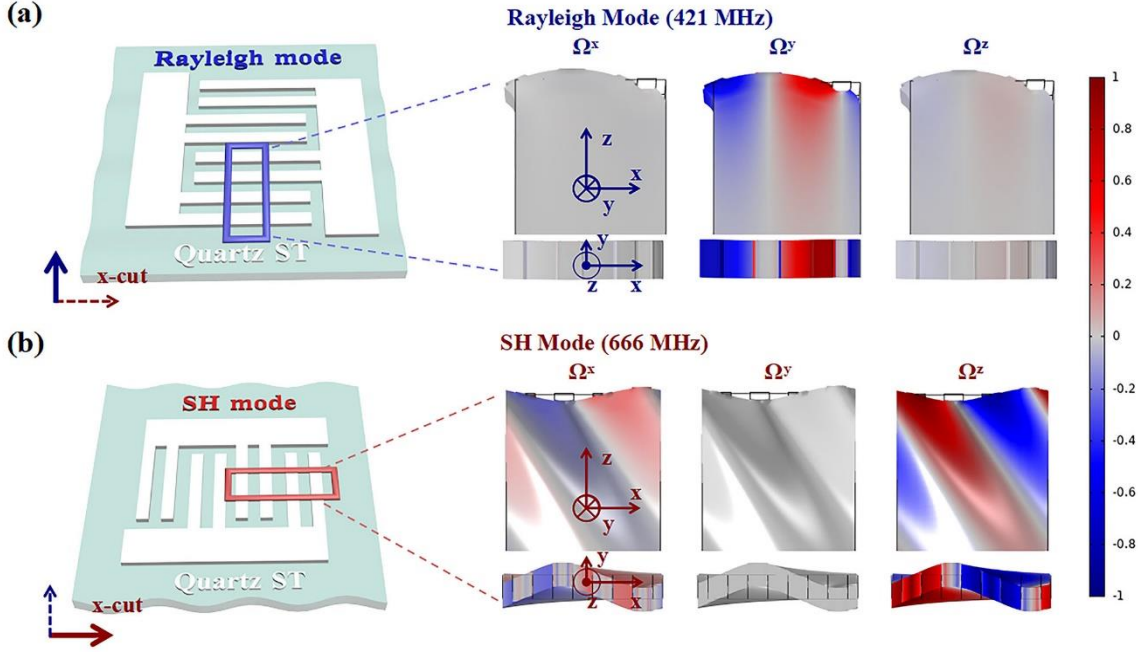

Fig. 1. FEM eigenfrequency simulations of the normalized $\Omega^x$, $\Omega^y$ and $\Omega^z$ for the R-SAW (a), and SH-SAW (b) with on a ST-cut quartz stacked with 50 nm-thick aluminum IDTs. The false colors represent the strength of vorticity, gray for weak vorticity, blue and red for large negative and positive vorticity, respectively. The solid arrow represents the main propagation direction of SAWs. Both R- and SH-SAWs can be excited by setting the direction of IDTs. The lattice deformations for SH- and R-SAW are also illustrated in the figure.

The propagation characteristics of R-SAW and SH-SAW on a ST-cut quartz were simulated by using a commercial COMSOL Multiphysics 5.6 software with piezoelectric and elastic dynamic modules. Split-finger interdigital transducers (IDTs) were designed along and perpendicular to the *X*-cut direction to generate SH- and R-SAWs, respectively. Periodic conditions are applied along the *x* and *y* directions, and the bottom of the substrate is fixed. Fig. 1(a) and 1(b) show the FEM eigenfrequency simulation results of the $\Omega^x$, $\Omega^y$ and $\Omega^z$ for the R- and SH-SAWs with the same wavelength of 7.5 μm. We assign that both SAWs propagate along the +*x* direction, and *z* is the normal direction of the plane. R-SAW can be regarded as coupled shear vertical wave and longitudinal wave. It has the displacement components along the *x* and *z* directions, i.e. $\boldsymbol{u} = (u_1, 0, i \cdot u_3)$, where the imaginary unit *i* represents the phase shift π/2 between two displacement components. Therefore, as shown in Fig. 1(a), R-SAW only has strong vorticity $\Omega_R{}^y$ along the *y* direction. For the SH-SAW in Fig. 1(b), although it only has the displacement component in the *y* direction, i.e $\boldsymbol{u} = (0, u_2, 0)$, it has strong vorticity $\Omega_{SH}{}^x$ and $\Omega_{SH}{}^z$ along the *x* and *z* directions, respectively. The values of vorticity generated by R- and SH-SAW can be determined by:

$$\Omega_R{}^y = \frac{\omega^2 u}{2c_t} \exp\left[-k_t z + i(kx - \omega t)\right]$$

$$\Omega_{SH}{}^x = \frac{\omega^2 u}{2c_t} \exp\left[-k_t z + i(kx - \omega t)\right] \quad (3)$$

$$\Omega_{SH}{}^z = i \cdot \frac{\omega^2 u}{2c_t} \exp\left[-k_t z + i(kx - \omega t)\right]$$

where $u$, $\omega$ and $c_t$ are the displacement amplitude, the angular frequency and the transverse velocity of the SAWs. For the R-SAW, the transverse wavenumber $k_t$ can be descripted by the wavenumber $k$: $k_t = k\sqrt{1 - \xi^2}$, where $\xi \approx (0.875 + 1.12\nu)/(1 + \nu)$ is a variable related to Poisson's ratio $\nu$. As for the SH-SAW, the following relation $k_t = k$ is satisfied. Because the phase velocity of SH mode (4995 m/s) is much larger than that of Rayleigh mode (3158 m/s) for a ST-cut quartz substrate, given the same wavelength of 7.5 μm, SH-SAW has a higher eigenfrequency (666 MHz) than that of R-SAW (421MHz), and therefore a greater amplitude of vorticity $\Omega$.

According to the SVC theory [24][27], the diffusion of spin accumulation $\delta\mu$ with respect to spatial and temporal variations can be expressed as:

$$\left(\partial_t - D\nabla^2 + \tau_{sf}^{-1}\right)\delta\mu = \hbar\partial_t\Omega \tag{4}$$

where $D$ is the diffusion constant and $\tau_{sf}$ represents the spin-flip time. Moreover, the spin current generated by spin accumulation can be expressed as[27]

$$J_s = \frac{\sigma_0}{e}\nabla\delta\mu \tag{5}$$

with conductivity $\sigma_0$ and elementary charge $e$. The polarization direction of spin currents (SCs) is related to the direction of effective vorticity generated by SAWs. Therefore, R-SAW generates the y-polarized SCs $J_s^Y$, while SH-SAW causes the x-polarized SCs $J_s^X$ and z-polarized SCs $J_s^Z$, all along the z-axis. By substituting Eq. (3) into Eq. (4), the expressions of the alternating SCs generated by both types of SAWs can be obtained. The SCs generated by R-SAW for $k_t z \ll 1$ can be described as [24][27]:

$$J_s^Y = \zeta J_s^{R'} e^{i(kx-\omega t)} \tag{6}$$

where

$$J_s^{R'} \approx \frac{\hbar\sigma_0\omega^3 u}{ec_t^2}\left(1+\frac{k_t^2\lambda_s^2}{1-\xi^2}\right)^{-1/4}\frac{\sqrt{1-\xi^2}}{\xi}\frac{d_{NM}}{\lambda_s} \tag{7}$$

As for the SH-SAW, the SCs can be given by:

$$\begin{aligned} J_s^X &= \zeta J_s^{SH'} e^{i(kx-\omega t)} \\ J_s^Z &= i\cdot\zeta J_s^{SH'} e^{i(kx-\omega t)} \end{aligned} \tag{8}$$

where

$$J_s^{SH'} \approx \frac{\hbar\sigma_0\omega^3 u}{ec_t^2}\left(1+k^2\lambda_s^2\right)^{-1/4}\frac{d_{NM}}{\lambda_s} \tag{9}$$

for $k_t z \ll 1$. Here, $\zeta$ is a normalization factor representing conversion efficiency between spin and mechanical rotation [27][28], $\lambda_s$ is the spin diffusion length satisfying $\lambda_s = \sqrt{D\tau_{sf}}$ [24], and $d_{NM}$ is the thickness of the NM layer. Clearly, the difference in the magnitudes of the SCs in Eq. (7) and (9) is caused by the different transverse wavenumbers of R-SAW and SH-SAW. Additionally, it can be seen from Eq. (7) and Eq. (9) that SC is also proportional to $\omega^3$. Due to the higher phase velocity, the eigenfrequency of SH mode is about 1.6 times that of Rayleigh mode for the same wavelength. Thus, SH-SAW can generate much stronger SC than that by R-SAW.

**B. The Effective Driven Field Induced by STT**

As illustrated in Fig. 2, the spin-polarized currents generated in the NM layer through SVC is injected into the FM layer along the thickness direction. Due to the STT effect, the magnetic moments in the FM layer will precess with the injection of SCs. The spin polarization directions of the SCs are different for two modes, thus resulting in different effective driven field $\mathbf{h}_{st}$, which can be written as follows:

$$\mathbf{h}_{st} = -\frac{\hbar T J_s}{2e\mu_0 M_s^2 d} \mathbf{a} \times \mathbf{m} \tag{10}$$

where $T$, $d$, $\mathbf{a}$ and $\mathbf{m}$ represent the spin transparency at the FM/NM interface, the thickness of FM, the unity spin polarization vector of $J_s$ and the unity magnetization vector, respectively[29][36][37].

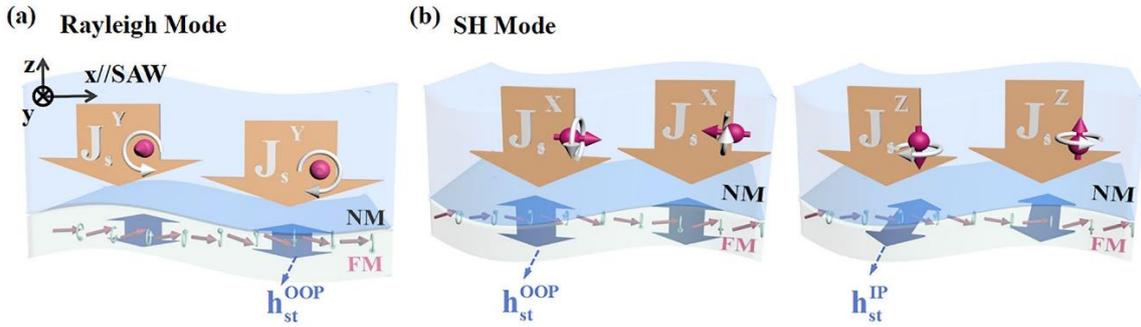

Fig. 2. Schematic illustration of SWR in the NM/FM structure excited by the R-SAW (a), and SH-SAW (b) via SVC. Both SAWs generate the alternating SCs along the z-axis via SVC. But the spin polarization directions are different for the two modes, $J_s^X$ and $J_s^Z$ for SH-SAW and $J_s^Y$ for R-SAW, thus resulting in different directions of STT-induced driven field $\mathbf{h}_{st}$.

By substituting Eq. (6) into Eq. (10), the effective driven field of R-SAW can be obtained as follows:

$$\mathbf{h}_{st}^{R} = \begin{pmatrix} h_{st}^{OOP} \\ h_{st}^{IP} \end{pmatrix} = \begin{pmatrix} h_1^R \\ h_2^R \end{pmatrix} = \begin{pmatrix} -\dfrac{\hbar T J_s^Y}{2e\mu_0 M_s^2 d}\cos(\varphi_0 - \varphi_G) \\ 0 \end{pmatrix} \quad (11)$$

where $\varphi_0$ ($\varphi_G$) is the angle between the equilibrium magnetization $\mathbf{m_0}$ (the x-axis) and the in-plane anisotropy hard-axis. Eq. (11) is solved in the 123-coordinate system with the 3-axis (1-axis) parallel to the equilibrium magnetization (the out-of-plane driven field $\mathbf{h}_{st}^{OOP}$) direction [18][34]. The setting of 123-coordinate system is shown in Fig. 3. Similarly, the effective driven field of the SH-SAW can be expressed as:

$$\mathbf{h}_{st}^{SH} = \begin{pmatrix} h_{st}^{OOP} \\ h_{st}^{IP} \end{pmatrix} = \begin{pmatrix} h_1^{SH} \\ h_2^{SH} \end{pmatrix} = \begin{pmatrix} \dfrac{\hbar T J_s^X}{2e\mu_0 M_s^2 d}\sin(\varphi_0 - \varphi_G) \\ -i \cdot \dfrac{\hbar T J_s^Z}{2e\mu_0 M_s^2 d} \end{pmatrix} \quad (12)$$

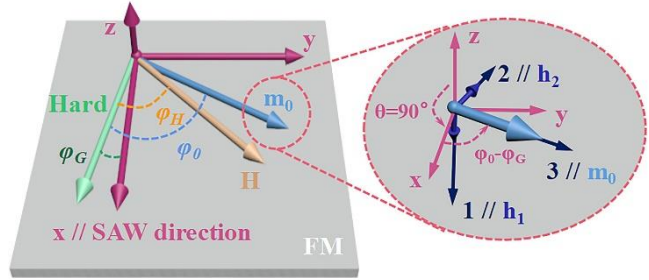

Fig. 3. The coordinate system setting for solving LLG equation. The inset shows relation between the two coordinate systems employed. The (x, y, z) coordinate system consists of the propagation direction of the SAW, the transverse in-plane direction, and the normal of the FM film. In the (1, 2, 3) coordinate system, which can be obtained by rotating the (x, y, z) coordinate system along the magenta dotted arrows, the 3-axis is parallel to the $\mathbf{m_0}$ direction, whereas the 1- and 2-axis parallel to the $\mathbf{h}_{st}^{OOP}$ and $\mathbf{h}_{st}^{IP}$ directions, respectively.

Thanks to its special vorticity component, the effective driven field excited by SH-

SAW has a non-zero in-plane component $h_2^{SH}$, which vanishes for R-SAW. In the FM film system with a strong demagnetization field, the in-plane component can excite the precession of the magnetic moments more effectively than the out-of-plane component. Therefore, the magnetic procession amplitude driven by SH-SAW is expected much larger than that by R-SAW.

**C. Analytical Model for SAW Power Absorption due to SWR**

If the frequency and wavenumber of SAWs match those of SWR, $\mathbf{h}_{st}$ can excite a resonantly enhanced magnetic precession, which results in the absorption of SAW power. Different from measurement via the inverse spin Hall effect, the energy dissipation from SWR is independent of the sign of torque, and the positive and negative spin currents would not be compensated. Next, we quantitatively compare the power absorption due to SWR for the two SAW modes, which reflects the energy conversion efficiency from SAWs to SWs. The absorbed power $P_{abs}$ can be described by [19][35]

$$P_{abs} = \frac{1}{2}\omega\mu_0 \int_{V_0} \text{Im}\left[\mathbf{h}_{st}^* \bar{\chi} \mathbf{h}_{st}\right] dV_0 \tag{13}$$

where $V_0$ and $\bar{\chi}$ represents the volume and the Polder susceptibility tensor of FM.

In section II.B, we have obtained the $\mathbf{h}_{st}$ of two SAW modes and the $\bar{\chi}$ can be obtained by solving the LLG equation, which is the spatial and temporal equation of motion for the magnetization $\mathbf{m}(x,t)$ under an effective magnetic field $\mathbf{H}_{eff}$ [18][33],

$$\frac{\partial \mathbf{m}(x,t)}{\partial t} = -\gamma \mathbf{m}(x,t) \times \mu_0 \mathbf{H}_{eff} + \alpha \mathbf{m}(x,t) \times \frac{\partial \mathbf{m}(x,t)}{\partial t} \tag{14}$$

where $\alpha$ is the Gilbert damping factor. The effective magnetic field $\mathbf{H}_{eff}$ comprises the external magnetic field $\mathbf{H}$, the in-plane uniaxial magnetic anisotropy field $\mathbf{H}_{ani}$, the magnetic shape anisotropy field $\mathbf{H}_k$, the effective dipolar field $\mathbf{H}_{dip}$, the magnetic exchange interaction $\mathbf{H}_{ex}$ and the effective driven field $\mathbf{h}_{st}$ induced by STT as follows:

$$\mu_0 \mathbf{H}_{\text{eff},123} = \mu_0 \mathbf{H} + \mu_0 \mathbf{H}_{ani} + \mu_0 \mathbf{H}_k + \mu_0 \mathbf{H}_{dip} + \mu_0 \mathbf{H}_{ex} + \mu_0 \mathbf{h}_{st}$$

$$= \mu_0 \mathbf{H}_{123} + \mu_0 H_{ani} (\mathbf{m}_{123} \cdot \mathbf{n}_{123}) \mathbf{n}_{123} + \mu_0 H_k \begin{pmatrix} m_1 \\ 0 \\ 0 \end{pmatrix}$$

$$- \mu_0 M_s \begin{pmatrix} G_0 m_1 \\ (1-G_0) m_2 \sin^2(\varphi_0 - \varphi_G) \\ 0 \end{pmatrix} - \frac{2A}{M_s} k^2 \begin{pmatrix} m_1 \\ m_2 \\ 0 \end{pmatrix} + \mu_0 \begin{pmatrix} h_1 \\ h_2 \\ 0 \end{pmatrix} \quad (15)$$

where $\mathbf{H}_{123}$, $\mathbf{m}_{123}$, and $\mathbf{n}_{123}$ are external magnetic field vector, unity magnetization vector and unity in-plane easy axis field vector in the 123-coordinate system, as shown in Fig. 3. Moreover, $H_{ani}$ represents the magnitude of the in-plane uniaxial anisotropic field. $H_k$ represents the out-of-plane surface anisotropic field caused by shape anisotropy, which can be given by $H_k = 2K_s / \mu_0 M_s d$, and $K_s$ is the constant of the surface perpendicular anisotropy. $M_s$ is the saturation magnetization, $A$ is the magnetic exchange stiffness, and $G_0 = (1 - e^{-|k|d}) / (|k|d)$ represents the dipolar spin wave term[34][35].

Thus, $\bar{\chi}$ can be solved by substituting Eq. (15) into Eq. (14) [18]:

$$\bar{\chi} = \frac{1}{C} \begin{pmatrix} \chi'_{11} & \chi'_{12} \\ \chi'_{21} & \chi'_{22} \end{pmatrix} \quad (16)$$

with

$$\chi'_{11} = H \cos(\varphi_0 - \varphi_H) - H_{ani} \cos(2\varphi_0) + M_s (1 - G_0) \sin^2(\varphi_0 - \varphi_G) + \frac{2A}{M_s} k^2 - \frac{i \omega \alpha}{\gamma}$$

$$\chi'_{12} = \chi'_{21} = -\frac{i\omega}{\gamma}$$

$$\chi'_{22} = H \cos(\varphi_0 - \varphi_H) - H_{ani} \cos^2 \varphi_0 - H_k + M_s G_0 + \frac{2A}{M_s} k^2 - \frac{i \omega \alpha}{\gamma} \quad (17)$$

$$C = \chi'_{11} \cdot \chi'_{22} - \chi'_{12} \cdot \chi'_{21}$$

And $\varphi_H$ is the angle of the external magnetic field $\mathbf{H}$ with respect to the in-plane anisotropy hard-axis.

Thus, for $kd \ll 1$, $P_{abs}$ can be approximately expressed as:

$$P_{abs} = P_{abs}^{st,OOP} + P_{abs}^{st,IP} + \text{cross terms}$$
$$\approx \frac{\gamma \mu_0 V_0}{2\alpha} \left[ \frac{\omega^2}{(\gamma M_s)^2} \left|h_{st}^{OOP}\right|^2 + \left|h_{st}^{IP}\right|^2 + \text{cross terms} \right] \quad (18)$$
$$\approx A_0^{st,OOP} \omega^8 + B_0^{st,IP} \omega^6 + \text{cross terms}$$

The in-plane and out-of-plane driven field components differ by one term of $\omega^2/(\gamma M_s)^2$ in their contributions to $P_{abs}$, indicating that the SWR driven by the out-of-plane field is inhibited by the demagnetizing field. Due to $\omega \ll \gamma M_s$, the contribution of the in-plane driven field component $h_{st}^{IP}$ is much greater than that of the out-of-plane component $h_{st}^{OOP}$. In addition, the lower SAW frequency, the larger difference between $P_{abs}^{st,OOP}$ and $P_{abs}^{st,IP}$ is expected. Moreover, the driven field component $h_{st}^{OOP}$ and $h_{st}^{IP}$ are proportional to $J_s$, which is proportional to $\omega^3$ (see Eq. (7) and Eq. (9)). Thus, the contributions of $P_{abs}^{st,OOP}$ and $P_{abs}^{st,IP}$ are proportional to the eighth and sixth order of frequency, respectively.

Next, we calculate the $P_{abs}$ excited by R- and SH-SAWs as a function of $\mathbf{H}$ for the same wavelength (7.5 μm) and displacement amplitude (0.1 nm) in a ST-cut quartz/Ni$_{81}$Fe$_{19}$ (20 nm) /Cu (200 nm) structure, as shown in Fig. 4(a) and 4(b), respectively. The $P_{abs}$ excited by SH-SAW is about 500000 times stronger than that by R-SAW. This benefits from the higher eigenfrequency of SH-mode, thus stronger spin current $J_s$ and STT-induced driven field $\mathbf{h}_{st}$. Moreover, compared with the out-of-plane component ($\mathbf{h}_{st}^{OOP}$) of R-SAW, the driven efficiency of the in-plane component ($h_{st}^{IP}$) of SH-SAW is also significantly improved. The detailed calculation parameters are given in Appendix A.

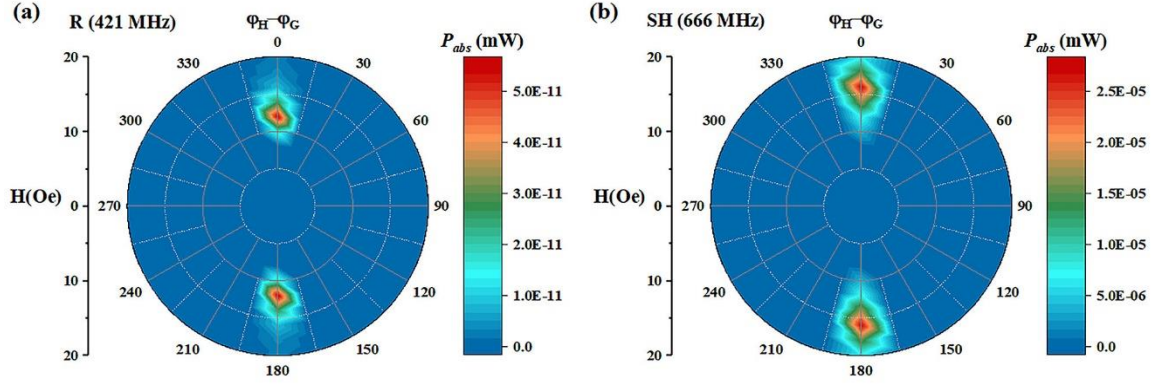

Fig. 4. Polar plot of the SWR power absorption $P_{abs}$ excited by R-SAW (a) and SH-SAW (b) as a function of external magnetic fields. Here, $\varphi_H - \varphi_G$ represents the angle between the applied magnetic field **H** and the SAW propagation direction.

## III. Experimental

To verify the theoretical predictions of SWR excited by SVC, we designed and fabricated split-finger IDTs on ST-cut quartz substrates along and perpendicular to the *X*-cut direction to generate SH- and R-SAWs, respectively (see Fig. 1). The split-finger design can reduce the reflection of SAWs, and is beneficial to obtain higher order harmonics. SH-SAW delay lines with wavelengths of 6, 7.5 and 12.5 μm and a R-SAW delay line with a wavelength of 7.5 μm were fabricated. Thus, SH-SAWs with center frequencies of 400, 666, 833 MHz and 1.20 GHz (the 3rd harmonic of 400 MHz) can be obtained. Meanwhile, R-SAWs with a fundamental frequency of 421 MHz and a 3rd harmonic frequency of 1.26 GHz were also obtained for comparison. The 50 nm-thick aluminum IDTs were prepared by magnetron sputtering with a 5 nm-thick titanium adhesion layer, and the spacing between two IDT pairs is 560 μm. In the spacing between the IDTs, a square-shaped $Ni_{81}Fe_{19}$ (20 nm)/Cu (200 nm) or $Ni_{81}Fe_{19}$ (20 nm) film of 500×500 μm² were deposited via sputtering and lithographically patterned. An in-situ external magnetic field of 250 Oe was applied during sputtering to induce an in-plane uniaxial magnetic anisotropy (IPUMA).

The transmission parameters $S_{21}$ of R- and SH-SAW delay lines were measured by a vector network analyzer (VNA, Agilent N5230A) using the setup illustrated in Fig. 5(a). The input microwave power was fixed at 10 dBm. The RF power absorbed by SWR can be normalized by

$$\Delta P^{\text{norm}}(H) = \frac{|P_{21}(H) - P_{21}(100\,\text{Oe})|}{P_{21}(100\,\text{Oe})} \quad (19)$$

where $P_{21}(H)$ represents the transmitted power calculated from $|S_{21}|$ at the SAW frequency under different external magnetic fields. $P_{21}(100\,\text{Oe})$ is the transmitted power measured at 100 Oe, which is sufficient to fully saturate $Ni_{81}Fe_{19}$. Fig. 5(b) plots the measured magnitude of the power transmission coefficient $|S_{21}|$ of the R-SAW and SH-SAW devices with the same wavelength of 7.5 μm, and the inset shows the optical photograph of the device.

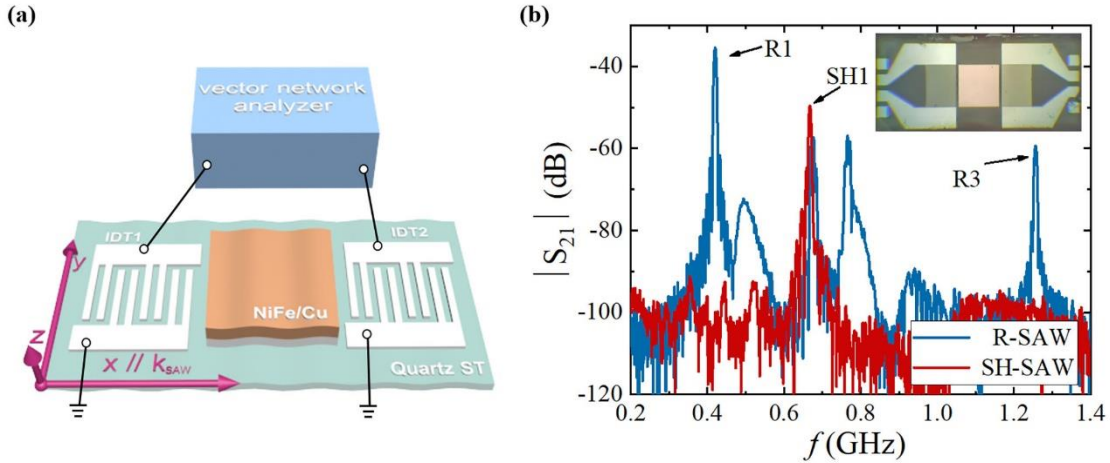

Fig. 5 (a) Schematic illustration of the setup for measuring $\Delta P^{\text{norm}}$ absorbed by spin wave resonance. (b) The measured frequency dependence of $|S_{21}|$ of R-SAW (blue line) and SH-SAW (red line) delay [lines with $\lambda = 7.5\,\mu m$. The inset shows the optical photograph of a SH-mode device.

## IV. Results and Discussion

In a $Ni_{81}Fe_{19}$ (20 nm) /Cu (200 nm) bilayer structure, SWR may also be directly excited in the FM layer by the Barnett field[25] and/or the magneto-rotation coupling[38], besides spin current injection from the Cu layer by SVC. In order to clarify this, the normalized power absorption $\Delta P^{norm}$ of two SH-SAW delayer lines are compared, one coated with a $Ni_{81}Fe_{19}$ single layer (denoted as D1), while the other with a $Ni_{81}Fe_{19}$/Cu bilayer (denoted as D2). Both devices have the same wavelength of 7.5 μm, thus the same fundamental frequency of 666 MHz. As shown in Fig. 6(a), $\Delta P^{norm}$ is almost zero for D1 coated with a $Ni_{81}Fe_{19}$e single layer. Therefore, the magnetic precession caused by either the Barnett field or the magneto-rotation coupling is very weak and can be ignored. However, a significant $\Delta P^{norm}$ about 3.67% was measured at about $\pm 14$ Oe for the D2 device coated with a $Ni_{81}Fe_{19}$/Cu bilayer, in spite of the small displacement of ST-cut quartz due to the low piezoelectric constant. Thus, the power absorption originates from the spin accumulation in the Cu layer via SVC rather than other magnon-phonon coupling effects.

In addition, a R-SAW delay line coated with a $Ni_{81}Fe_{19}$/Cu bilayer (denoted as D3) was also fabricated and tested for comparison. Fig. 6(b) shows the $\Delta P^{norm}$ of the device measured at 421 MHz and 1.26 GHz (the 3rd harmonic) upon applying external magnetic fields along the SAW propagation direction. Almost no $\Delta P^{norm}$ is observed at 421 MHz, meaning that the power absorption at low frequencies is close to the noise level, which is consistent with our analysis in Fig. 41. $\Delta P^{norm}$ only reaches a maximum value of 0.47% at a magnetic field of $\pm 8$ Oe for the 3rd R mode at 1.26 GHz. This can be understood by the highly nonlinear frequency dependence of $\Delta P^{norm}$. Previous work has shown that the maximum of the $\Delta P^{norm}$ of R-SAW is proportional to $f^7$ [30]. But this $\Delta P^{norm}$ value is still much less than that of D2. Notice that the D3 device has the same wavelength as that of D2. Considering the measured $\Delta P^{norm}$ and the wavelength difference between SH-SAW (666 MHz) and R-SAW (1.26 GHz), the power absorption of SH-SAW is about four orders of magnitudes greater than that of R-SAW at the same wavelength of 7.5 μm. These results clearly prove that SH-SAW is much more efficient than R-SAW to induce SWR, which is in good accordance with our theoretical calculation in Sec. II C.

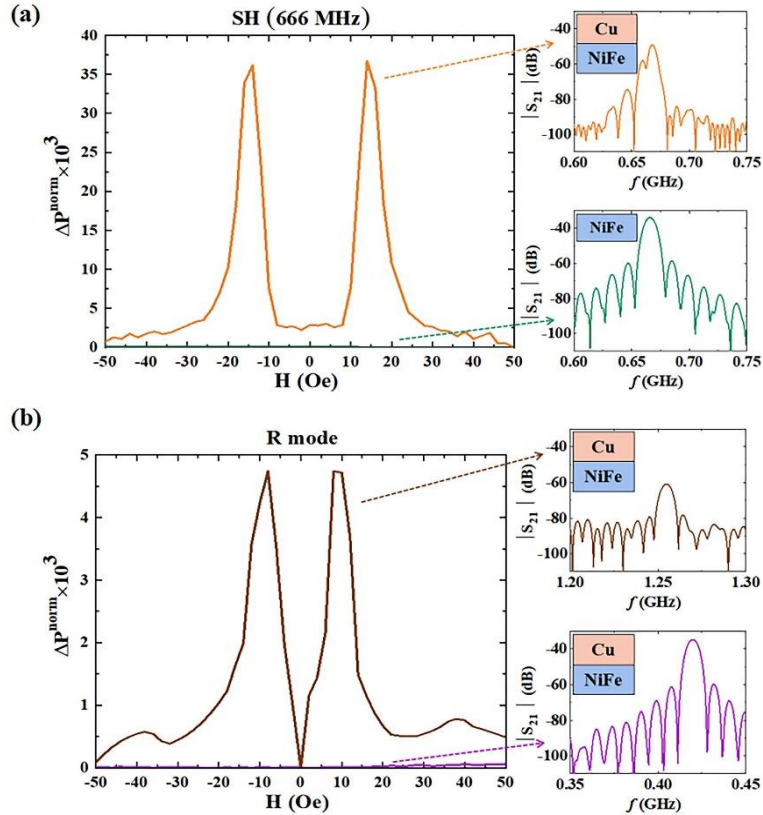

Fig. 6 Field dependent normalized power absorption of (a) the D1 and D2 SH-SAW delay lines measured at 666 MHz, and (b) the D3 Rayleigh-SAW delay line measured at 421 MHz and 1.26 GHz (the 3rd harmonics). All tests were performed with the external magnetic field along the SAW propagation direction. The right-hand insets plot the $|S_{21}|$ measured at 100 Oe after time-domain gating. The detail of time-domain gating method is provided in Appendix B.

Next, using the input power, the measured $|S_{21}|$ and the $\Delta P^{norm}$ at $\varphi_H - \varphi_G = 0$ in Fig. 6, the displacement amplitudes $u$ corresponding to different acoustic modes are obtained by solving Eq. (16), 0.043 nm for the SH-SAW at 666 MHz and 0.039 nm for the R-SAW at 1.26 GHz. Then, these two values were substituted into Eq. (13)-(16) to further calculate the field-dependent $\Delta P^{norm}$ at different angles of $\varphi_H - \varphi_G$, as shown in Fig. 7. The detailed calculation parameters are given in Appendix A. Fig. 7(a) and (b) also plot the experimental results. It can be seen that the experimental results are in good agreement with the theoretical fitting ones in terms of both angle and magnitude dependences. Moreover, the polar plot of the field dependent $\Delta P^{norm}$ show that the strongest absorption locates at $\varphi_H - \varphi_G = 0°$, but completely vanishes close to 90°. This angular dependence is also an important criterion for distinguishing SVC from other types of magnon-phonon couplings [27][30].

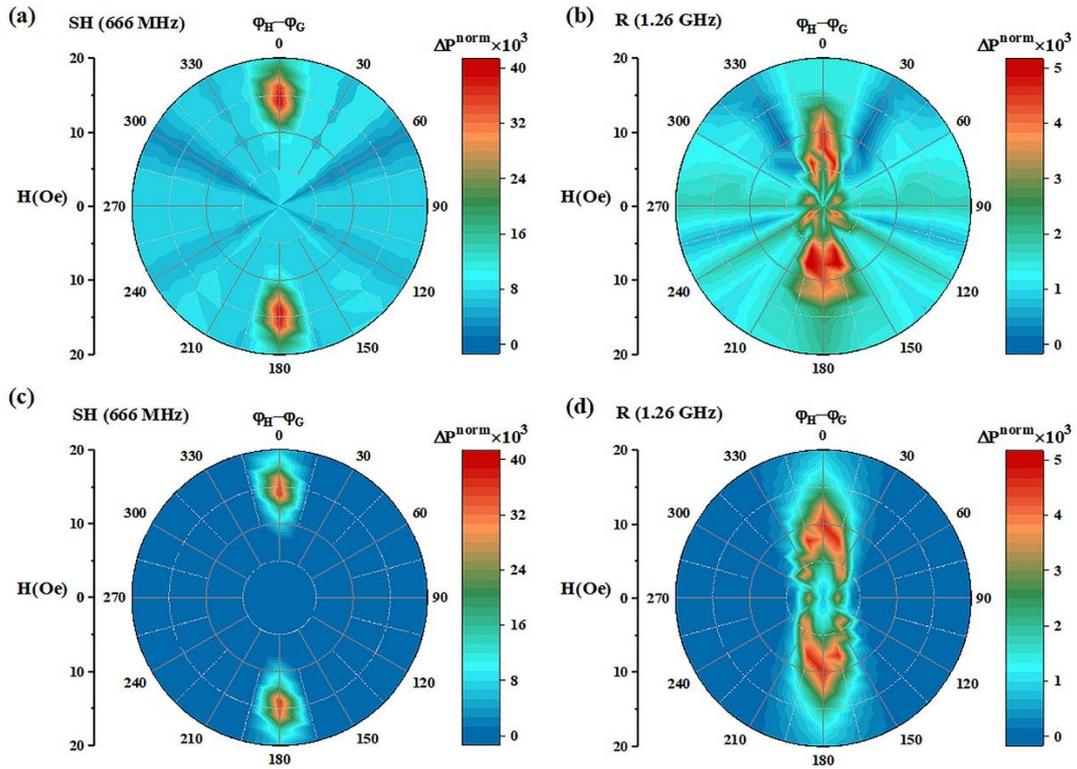

Fig. 7 Polar plot of measured (a and b) and calculated (c and d) field-dependent normalized power absorption of the SH-SAW delay line at 666 MHz (a and c) and Rayleigh-SAW delay line at 1.26 GHz

(b and d).

Finally, we measured all three SH-SAW delay lines with wavelengths of 6, 7.5 and 12.5 μm, and the $\Delta P^{norm}$ at 400, 666, 833 and 1200 MHz (the 3rd harmonic of 400 MHz) are plotted in Fig. 8(a). With the increase of frequency, the resonance magnetic field gradually increases, which is consistent with the dispersion relation of SWs [17]. Meanwhile, $\Delta P^{norm}$ dramatically increases to 28.5% at 1200 MHz, much larger than that excited by R-SAW at 1.26 GHz (0.47%) in Fig. 6(b). It is worth noting this value means a very high SAW-SW conversion efficiency, close to those based on the magnetoelastic coupling [17][30]. Furthermore, Fig 8(b) shows the *ln-ln* plot of the frequency-dependent $\Delta P^{norm}$ of SH-SAW delay lines, where $f_0$ is set to 400 MHz. The slop of the linear fitting is 3.79 ± 0.14, indicating the highly nonlinear frequency dependence of SVC excited by SH-SAW. Theoretically, $\Delta P^{norm}$ can be obtained by $P_{abs}/P_{SAW}$, where $P_{SAW}$ is the transmission power of SAWs. $P_{SAW}$ satisfies the relation of $P_{SAW} = \omega F_0 W u^2$ [39], where $F_0$ and $W$ represent the approximately numerical calculated constant and the finger length of the IDT, respectively. Combined with Eq. (18), the maximal $\Delta P^{norm}$ due to SWR is expected to be proportional to $f^5$ for SH-SAW, due to the dominated in-plane driven field component. The reason of this declined frequency dependence of $\Delta P^{norm}$ in our experiments may be related to the increase of damping factor of $Ni_{81}Fe_{19}$. As shown in the inset of Fig. 8(b), the SWR linewidth deviates from the typical linear relationship with frequency, indicating that the effective damping factor is not a constant. A higher frequency corresponds to a higher effective damping factor, thus lower $\Delta P^{norm}$ according to Eq. (18). Further studies are needed to determine the source of this frequency-dependent damping factor for SVC driven

SWR, however, this is outside the scope of current work.

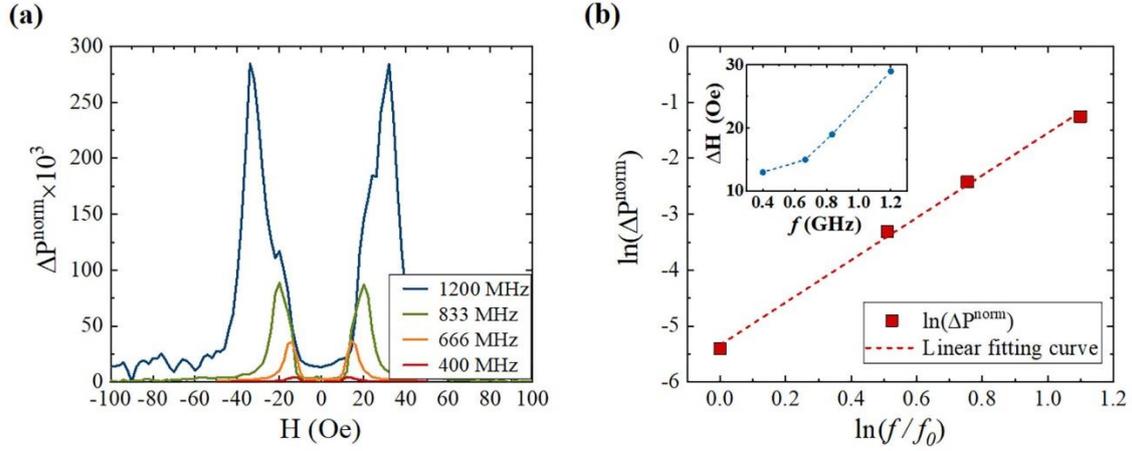

Fig. 8 (a) Field dependent normalized power absorption of SH-SAWs measured at different frequencies, where the external magnetic field is applied along the SAW propagation direction ($\varphi_H - \varphi_G = 0°$). (b) The frequency dependent normalized power absorption of SH-SAWs. The red dashed line is the linear fitting curve and $f_0$ is set to 400 MHz. The inset show the variation of the SWR linewidth $\Delta H$ with frequency.

## V. Conclusion

In summary, we present the theoretical model of spin wave resonance in the FM/NM bilayer excited by two kinds of SAWs via SVC. This model can describe the measured field- and angle- dependent power absorption of R- and SH-SAW delay lines very well. A four orders of magnitudes stronger power absorption has been demonstrated in SH-SAWs, compared to that in R-SAWs with the same wavelength, which manifests the high energy conversion efficiency from SH-SAW to SW. This can be attributed to its high phase velocity thus high eigenfrequency, distinct vorticities and the effective in-plane driven field. In addition, we also observed a high-order frequency dependence of normalized power absorption, indicating that the SWR excited by SH-SAW via SVC can be comparable to MEC at GHz frequencies. Our results pave the way to apply the spin-vorticity coupling to excite magnons in solid-state spintronic devices.

**Acknowledgement**

This work is supported by the National Natural Science Foundation of China (Grant No. 61871081) and the Natural Science Foundation of Sichuan Province under Grant No. 2022NSFSC0040.

**Appendix A: Theoretical calculation parameters**

Table □ shows the parameters used in the theoretical calculation in Fig. 4 and Fig. 7 (c) (d).

Table □. Parameters used in the theoretical calculations

| Parameter | Value |
|---|---|
| the saturation magnetization of $Ni_{81}Fe_{19}$, $M_s$ (kGs) | 9.8 |
| the gyromagnetic ratio of $Ni_{81}Fe_{19}$, $\gamma$ (MHz/Oe) | 17.6 |
| the Gilbert damping factor of $Ni_{81}Fe_{19}$, $\alpha$ | 0.01 |
| thickness of the Cu layer, $d_{NM}$ (nm) | 200 |
| thickness of $Ni_{81}Fe_{19}$, $d$ (nm) | 20 |
| length and width of $Ni_{81}Fe_{19}$ (μm) | 500, 500 |
| conductivity of Cu, $\sigma_0$ (S/m) | $5.8 \times 10^7$ |
| the reduced Planck constant, $\hbar$ | $1.05 \times 10^{-34}$ |
| elementary charge, $e$ (C) | $1.6 \times 10^{-19}$ |
| the transverse sound velocity of Cu, $c_t$ (m/s) | 2270 |
| the Poisson ratio of Cu, $\upsilon$ | 0.343 |
| the spin diffusion length of Cu, $\lambda_s$ (nm) | 350 |
| the normalization factor, $\zeta$ | $2.328 \times 10^8$ |
| the spin transparency at the NiFe/Cu interface, $T$ | 0.074 |
| the constant of the surface perpendicular anisotropy, $K_s$ (J/m$^2$) | $6 \times 10^{-4}$ |
| the magnetic exchange stiffness, $A$ (J/m$^3$) | $9.5 \times 10^{-12}$ |
| the in-plane uniaxial magnetic anisotropy field of $Ni_{81}Fe_{19}$, $H_{ani}$ (Oe) | 10 (Fig. 4(a) )<br>10 (Fig. 4(b) )<br>9 (Fig. 7(c) )<br>11 (Fig. 7(d) ) |
| the displacement amplitude, $u$ (nm) | 0.1 (Fig. 4(a) )<br>0.1 (Fig. 4(b) )<br>0.043 (Fig. 7(c) )<br>0.039 (Fig. 7(d) ) |

**Appendix B: Time-domain gating method**

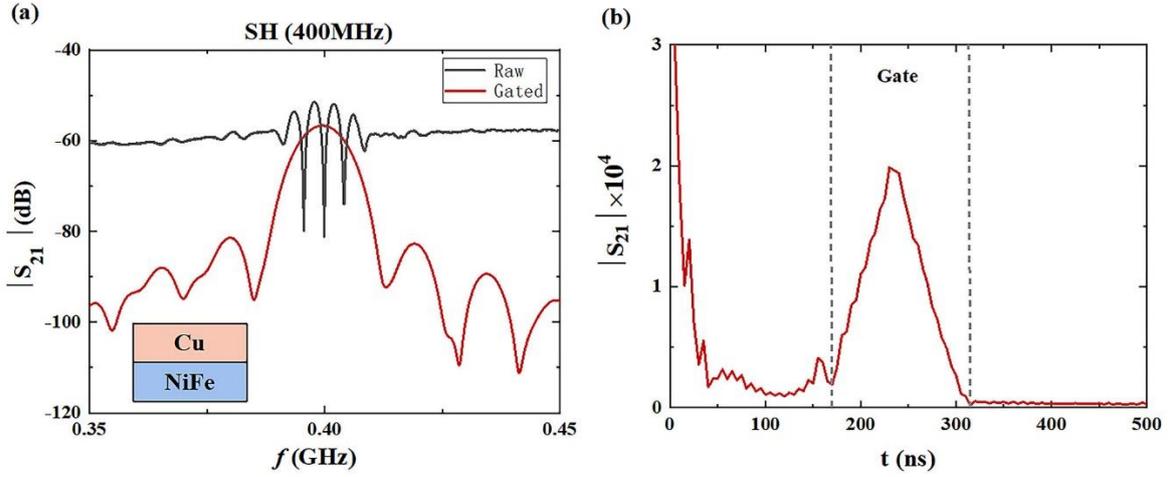

Fig. 9 (a) The frequency-domain transmission parameter $|S_{21}|$ of the D2 delay line at 400MHz before (gray line) and after (red line) the time domain gating. (b) The time-domain transmission parameter $|S_{21}|$ of the original data shown in Fig 9(a), where the area surrounded by the dotted line is the time-domain gating region.

The signal transmitted by IDT will be accompanied by interference of electromagnetic wave. Due to the large difference in transmission speed between SAWs and EMWs, the interference of EMWs can be effectively eliminated by the time-domain gating method, as shown in Fig. 9.